\documentclass[prd,article,nofootinbib,twocolumn,preprintnumbers,amsmath,amssymb,aps]{revtex4-1}
\usepackage{hyperref}
\usepackage{graphicx,epstopdf,amsmath,amsfonts,amssymb,appendix,comment} 
\usepackage{color,slashed,subfigure,setspace,footnote,multirow,longtable,braket}
\usepackage{epsfig,mathrsfs,latexsym,color,url,etoolbox}
\usepackage{bbm}
\usepackage[letterpaper, portrait, margin=0.65in]{geometry}
\definecolor{nicered}{rgb}{0.7,0.1,0.1}
\definecolor{nicegreen}{rgb}{0.1,0.5,0.1}

\usepackage{graphicx}
\usepackage{dcolumn}
\usepackage{bm}
\usepackage[table]{xcolor}
\usepackage{multirow}
\usepackage{comment}
\usepackage{url}
\usepackage{tcolorbox}
\usepackage[T1]{fontenc}

\definecolor{kjkblue}{rgb}{0.39, 0.589, 0.6914}

\DeclareMathAlphabet{\mathpzc}{OT1}{pzc}{m}{it}

\begin{document}

\preprint{FERMILAB-PUB-19-541-T}

\title{Prospects of Measuring Oscillated Decay-at-Rest Neutrinos at Long Baselines}

\author{Roni Harnik}
\author{Kevin J. Kelly}
\author{Pedro A.N. Machado}
\affiliation{Theoretical Physics Department, Fermilab, P.O. Box 500, Batavia, IL 60510, USA}

\date{\today}

\begin{abstract}
In addition to the next generation of beam-based neutrino experiments and their associated detectors, a number of intense, low-energy neutrino production sources from decays at rest will be in operation. In this work, we explore the physics opportunities with decay-at-rest neutrinos for complementary measurements of oscillation parameters at long baselines.
The J-PARC Spallation Neutron Source, for example, will generate neutrinos from a variety of decay-at-rest (DAR) processes, specifically those of pions, muons, and kaons. Other proposed sources will produce large numbers of stopped pions and muons. We demonstrate the ability of the upcoming Hyper-Kamiokande experiment to detect the monochromatic kaon decay-at-rest neutrinos from J-PARC after they have travelled several hundred kilometers and undergone oscillations. This measurement will serve as a valuable cross-check in constraining our understanding of neutrino oscillations in a new regime of neutrino energy and baseline length. We also study the expected event rates from pion and muon DAR neutrinos in liquid Argon and water detectors and their sensitivities to to the CP violating phase $\delta_\mathrm{CP}$.
\end{abstract}

\maketitle

\section{Introduction}

The discovery that neutrinos oscillate, and therefore have mass, has revolutionized our understanding of the lepton sector of the standard model of particle physics. Since this discovery, several generations of experiments have been designed, been built, and collected data to better understand the rich phenomenon of neutrino oscillations. Future experiments~\cite{Abi:2018dnh,Abe:2011ts,An:2015jdp} are on the horizon to pin down the few remaining unknowns associated with standard paradigm of neutrino oscillations. 

\begin{figure}
    \centering
    \includegraphics[width=\linewidth]{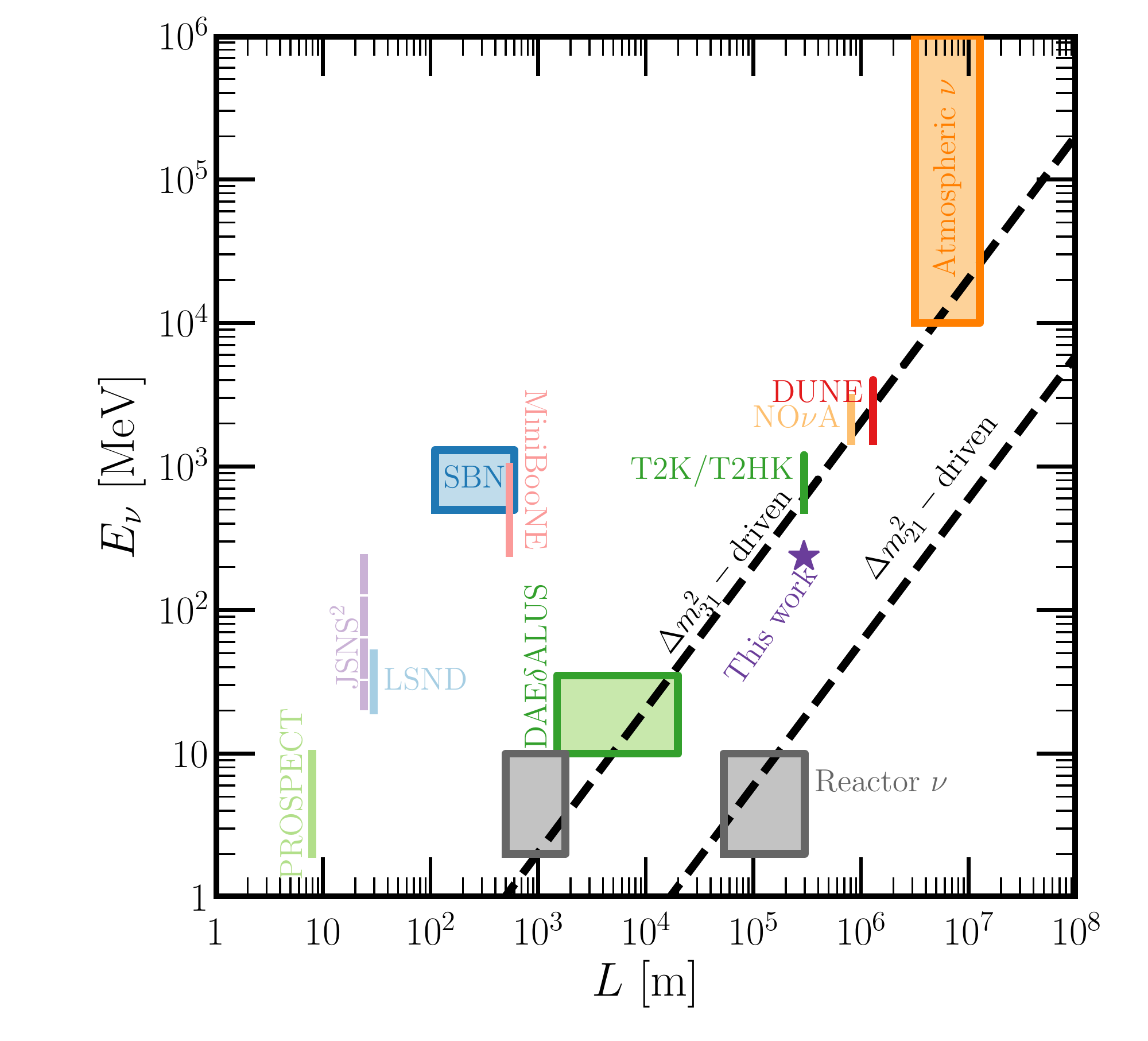}
    \caption{Baseline lengths and typical neutrino energies for a variety of searches for neutrino oscillations. The purple star labelled ``This work'' corresponds to neutrinos coming from kaon decay-at-rest travelling the distance between JSNS and Hyper-Kamiokande, 295 km (see Sec.~\ref{sec:ExpDetails}). The proposals of Sec.~\ref{sec:other} using $\pi$DAR and $\mu$DAR overlap with the region probed by DAE$\delta$ALUS.}
    \label{fig:LsEs}
\end{figure}

While this paradigm fits the vast majority of data exceptionally well, a few unexplained experimental results persist. The standard picture predicts that two ``frequencies,'' governed by two neutrino mass-squared splittings, drive all oscillations, however, evidence exists for a third, larger mass-squared splitting that would indicate a beyond-the-standard-model fermion existing and mixing with the light neutrinos (see e.g. Refs~\cite{Gariazzo:2017fdh, Dentler:2018sju, Diaz:2019fwt}). In order to explore these hints, and to over-constrain our knowledge of neutrino mixing, it is imperative to measure oscillations in as many domains as possible. When treating neutrino oscillations in vacuum, a given oscillation probability is determined by the ratio of the distance travelled by the neutrino to its energy. We argue that measuring oscillation probabilities for as many unique distances and energies is a compelling way to over-constrain the standard three-massive-neutrinos paradigm. Existing and planned measurments of oscilation probabilities, as well as those we propose here, are shown in Fig.~\ref{fig:LsEs} in the plane of neutrino energy and baseline. 

One specific source of neutrinos that has not been explored fully is that coming from processes in which mesons and muons decay at rest (DAR) into neutrinos. When this is a two-body decay, the neutrino flux is monoenergetic, and if oscillations associated with these neutrinos are observed, then the baseline length and neutrino energy of this process can be determined nearly perfectly (see the purple star in Fig.~\ref{fig:LsEs}). Neutrinos emerging from muon-, pion-, and kaon-decay-at-rest have been considered~\cite{Spitz:2012gp,Spitz:2014hwa} and observed~\cite{Armbruster:2002mp, Aguilar-Arevalo:2018ylq,Aguilar:2001ty}, however -- except for an oscillation-related interpretation of the anomalous muon-decay-at-rest result of the LSND experiment~\cite{Aguilar:2001ty} -- these neutrinos have not previously been measured after undergoing oscillations. In this work, we present a potential capability within oscillations from the three-massive-neutrinos paradigm: neutrinos from kaon-decay-at-rest produced in the J-PARC Spallation Neutron Source (JSNS)~\cite{Ajimura:2017fld}, traveling several hundred kilometers to Hyper-Kamiokande (HK) and interacting in a water cerenkov detector~\cite{Abe:2011ts}.
While this measurement does not provide any more powerful information if the three-massive-neutrinos paradigm is assumed to be true, it provides a consistency check on both current and future measurements. A measurement of this nature can test for other deviations from three-neutrino predictions, such as sterile neutrinos, non-standard neutrino interactions (see Ref.~\cite{Dev:2019anc} for a recent review), and other scenarios.

We also explore other opportunities for measuring DAR neutrinos at long baselines to test and measure the three neutrino framework. In particular we consider $\mu$DAR and $\pi$DAR fluxes, the measurement of which, however, would require either new intense sources or new detectors. 
To this end the DAE$\delta$ALUS~\cite{Conrad:2009mh} experiment proposes to use compact, but intense, cyclotron proton sources to generate a large number of $\pi$DAR neutrinos and their detection in a large water cerenkov detector such as Hyper-K over a variety of baselines in the 1-20~km range.  
In this work we briefly study the event rates of $\pi$DAR neutrinos in a large liquid Argon (LAr) detector such as DUNE, as well as the rate of inverse beta decay events of $\mu$DAR anti-neutrinos in large scintillator or water detectors. Like DAE$\delta$ALUS, these rates are sensitive to the CP violating phase in the three neutrino framework and can thus be complementary to measurements in neutrino beams.
 
This manuscript is organized as follows: in Section~\ref{sec:Oscillations} we discuss the phenomenon of neutrino oscillations, specifically focusing on the low energy, long baseline setup we are interested in. In Section~\ref{sec:ExpDetails}, we provide the relevant experimental details about JSNS and HK assumed in this work, as well as the strategy for detection of this process and reduction of background processes. 
In Section~\ref{sec:other} we consider other opportunities to test the three neutrino paradigm with DAR sources.
In Section~\ref{sec:Conclusions}, we offer some concluding remarks.

\section{Low-Energy, Long Distance Oscillations}\label{sec:Oscillations}
Neutrino oscillations are governed by a characteristic scale, determined by the mass-squared splitting $\Delta m_{ji}^2 \equiv m_j^2 - m_i^2$. A great deal of experimental evidence suggests that there are two non-zero mass-squared splittings, commonly referred to as the ``atmospheric mass splitting'' $\Delta m_{31}^2 \approx 2.5\times 10^{-3}$ eV$^2$  \cite{Abe:2017vif, Aartsen:2017nmd, NOvA:2018gge} and the ``solar mass splitting''\footnote{Current data from reactor (KamLAND) and solar experiments disagree on this parameter at the 2$\sigma$ level - this tension could be resolved or accelerated in the next generation of experiments, specifically JUNO and DUNE~\cite{Capozzi:2018dat}.} $\Delta m_{21}^2 \approx 7 \times 10^{-5}$ eV$^2$ \cite{Abe:2010hy, Aharmim:2011vm, Gando:2013nba, Agostini:2017cav, Agostini:2018uly}. Anomalous experimental results -- among them the LSND, MiniBooNE, and short-baseline reactor antineutrino experiments -- provide hints of a new mass splitting $\Delta m^2 \sim 1$ eV$^2$ \cite{Aguilar:2001ty, Ko:2016owz, Alekseev:2018efk, Aguilar-Arevalo:2018gpe}.

Understandably, a significant number of oscillation experiments have focused on ranges of baseline length $L$ and neutrino energy $E_\nu$ for which oscillations due to these three splittings are most significant. The phase that governs oscillation physics (in vacuum\footnote{For the energies and baselines of interest in this work, matter effects are not important in modifying neutrino oscillation probabilities.}) is
\begin{equation}
    P_{\alpha\beta} \equiv P(\nu_\alpha \to \nu_\beta) \propto \sin^2{\left( \frac{\Delta_{ij}}{2}\right)},
\end{equation}
where $\Delta_{ij} \equiv \Delta m_{ij}^2 L/2E_\nu$. When using experimentally-suitable units,
\begin{align}
    \Delta_{ij} &=& 2.534 \left(\frac{\Delta m_{ij}^2}{1\ \mathrm{eV}^2}\right) \left(\frac{L}{1\ \mathrm{km}}\right) \left(\frac{1\ \mathrm{GeV}}{E_\nu}\right) \\
    &=& 2.534 \left(\frac{\Delta m_{ij}^2}{1\ \mathrm{eV}^2}\right) \left(\frac{L}{1\ \mathrm{m}}\right) \left(\frac{1\ \mathrm{MeV}}{E_\nu}\right).
\end{align}
The effects of oscillations will be maximized when $\Delta_{ij}$ is an odd multiple of $\pi$.

\begin{figure}[!t]
\begin{center}
\includegraphics[width=\linewidth]{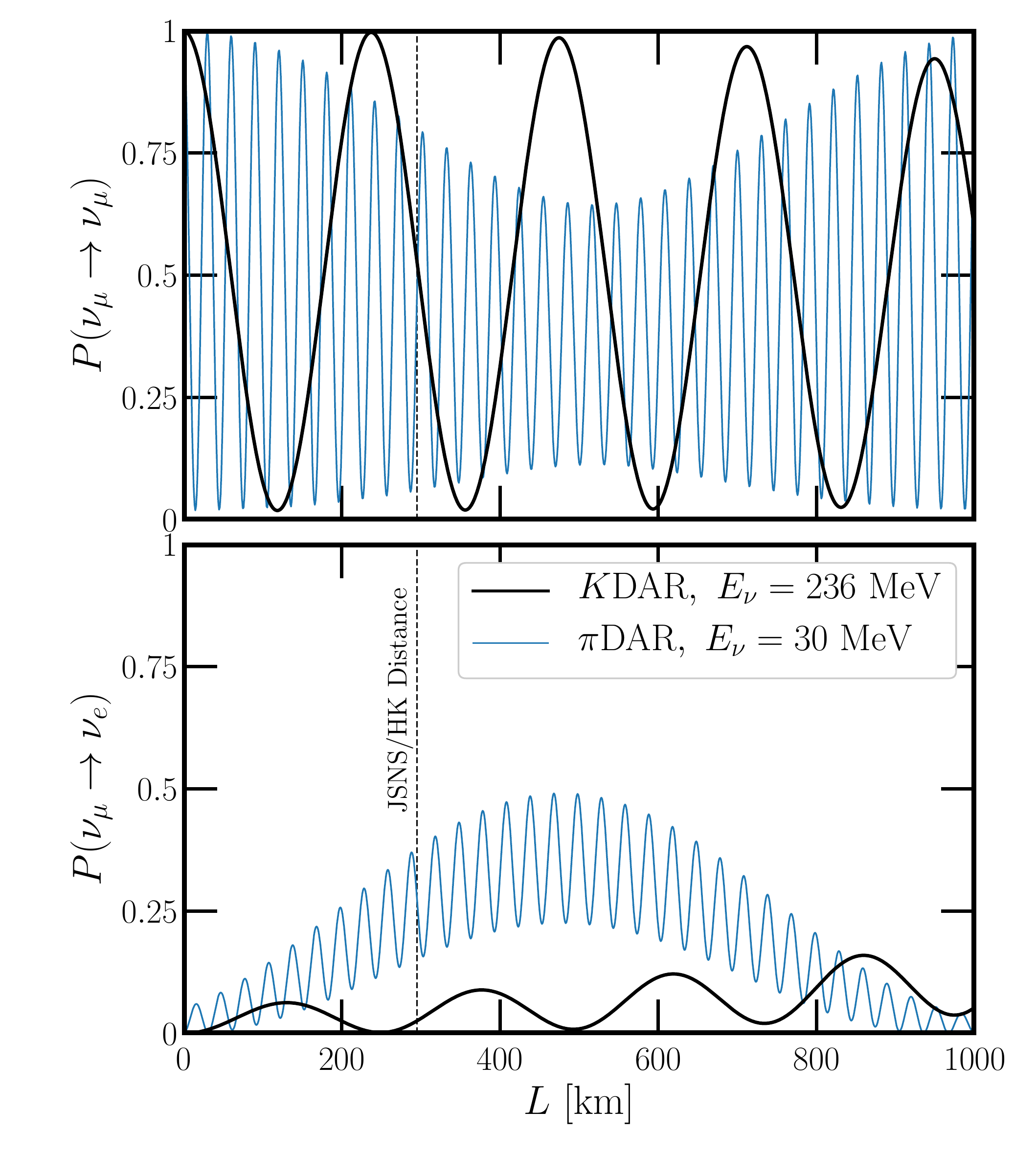}
\caption{Oscillation probabilities as a function of baseline length $L$ for $P(\nu_\mu \to \nu_\mu)$ (top) and $P(\nu_\mu \to \nu_e)$ (bottom) for neutrinos coming from kaon decay-at-rest (black) and pion decay-at-rest (blue). \label{fig:Probs}}
\end{center}
\end{figure}

Fig.~\ref{fig:LsEs} displays a subset of current and future neutrino oscillation experiments. Experiments sensitive to the atmospheric mass-squared splitting populate the region of this parameter space where $L/$km $\sim 500 E_\nu/\mathrm{GeV}$, where the first oscillation maximum occurs (the line labelled ``$\Delta m_{31}^2$-driven'' in Fig.~\ref{fig:LsEs}). Experiments related to the $1$ eV$^2$ sterile neutrino anomalies populate the top-left region of this parameter space, where $L$/km $\sim 1.25k (E_\nu/\mathrm{GeV})$ ($k$ is an odd integer). See Refs.~\cite{Diaz:2019fwt,Boser:2019rta} for a detailed survey of these experiments and others not shown in Fig.~\ref{fig:LsEs}. The JSNS$^2$ experiment is an upcoming experiment that satisfies this length/energy relation and will test the eV$^2$ sterile anomalies~\cite{Ajimura:2017fld}. If the neutrinos produced for this experiment travel a longer baseline, such as the distance from JSNS to Hyper-Kamiokande, oscillations will be governed by the atmospheric and solar mass splittings.

Notably, the majority of existing and future experiments rely on measuring the neutrino oscillation probability over some finite range of energies. The upcoming Deep Underground Neutrino Experiment (DUNE)~\cite{Abi:2018dnh}, for instance, aims to have a broad band beam, allowing for measurements across many different neutrino energies. Even experiments that aim to have a narrow beam energy, such as the NuMI Off-axis $\nu_e$ Appearance (NOvA) experiment~\cite{NOvA:2018gge}, still have some spread in neutrino energy (NOvA, for instance, has the bulk of its neutrino events between approximately $1-3$ GeV). In contrast, if neutrinos from a two-body decay-at-rest process are detectable, the energy is fixed due to kinematics and therefore it is known. Combined with a precise knowledge of the source-detector distance, this gives a measurement of the neutrino oscillation probability at a very specific combination of $L$ and $E_\nu$. Combining several such measurements  across many different but known $L$ and $E_\nu$ can be regarded as precision neutrino spectroscopy, which in turn would contribute to completing our understanding of neutrino oscillations.

A neutrino emerging from a two-body DAR process, where a charged meson $\mathfrak{m^\pm}$ decays into a charged lepton $\ell^\pm$ and a neutrino (or antineutrino) has a fixed energy,
\begin{equation}
E_\nu = \frac{m_\mathfrak{m}^2 - m_\ell^2}{2m_\mathfrak{m}},
\end{equation}
where $m_\mathfrak{m}$ is the decaying meson mass and $m_\ell$ is the charged lepton mass. We will be interested in the decays of $\pi^\pm$ and $K^\pm$, which decay predominantly via $\mathfrak{m^\pm} \to \mu^\pm \nu$. The energies of the neutrinos in these processes are $E_\nu = 29.8$ MeV ($\pi$DAR) and $E_\nu = 236$ MeV ($K$DAR).

Under the three-massive-neutrinos paradigm, we may calculate the oscillation probability of a $\pi$DAR or $K$DAR neutrino that is emitted as a $\nu_\mu$ and travels a distance $L$. These are too low-energy to produce charged $\tau$ leptons, and therefore we focus on the oscillation probabilities $P(\nu_\mu\to \nu_\mu)$ and $P(\nu_\mu \to \nu_e)$. The oscillation probabilities are shown in Fig.~\ref{fig:Probs}, where the black (blue) lines indicate oscillation probabilities for $K$DAR ($\pi$DAR) neutrinos. The $\pi$DAR neutrinos are too low-energy to produce charged $\mu$ leptons, and the cross section for these low energy neutrinos to produce $e^-$ in a charged current interaction on water are too low to detect, as we will discuss in the next section. However, if the $\pi$DAR oscillation probability could be measured at $L \approx \mathcal{O}(100\ \mathrm{km})$, perhaps using a different target than water, such a measurement would be sensitive to many effects of the three-massive-neutrinos formalism. At these distances, as evident in Fig.~\ref{fig:Probs}, the $\pi$DAR neutrino oscillation probability is driven by two frequencies (and their interference) -- the atmospheric and solar mass-squared splittings are both relevant for these energies and distances. Because the interference of these two is relevant, we would also be sensitive to the CP-violating phase $\delta_\mathrm{CP}$, the least well-understood component of the lepton mixing matrix.

The $K$DAR oscillation probability in Fig.~\ref{fig:Probs} (black line), however, is driven predominantly by the atmospheric mass-squared splitting. This is apparent in Fig.~\ref{fig:LsEs}: the star labelled ``this work'' corresponds to the $K$DAR neutrino energy and a baseline of $295$ km, where oscillations are relatively insensitive to $\Delta m_{21}^2$. As $L\to 1000$ km, we see that the effects of $\Delta m_{21}^2$ are starting to be important in the bottom panel of Fig.~\ref{fig:Probs}, however, even then, we have not reached the first maximum of the oscillations driven by $\Delta m_{21}^2$. In this regime, ignoring subdominant effects due to matter interactions, we may express the oscillation probabilities as
\begin{eqnarray}
P(\nu_\mu\to\nu_\mu) &\approx& 1 - 4|U_{\mu 3}|^2 (1 - |U_{\mu 3}|^2) \sin^2{\left(\frac{\Delta m_{31}^2 L}{4E_\nu}\right)} \label{eq:Pmumu}\\
P(\nu_\mu\to\nu_e) &\approx& 4|U_{\mu 3}|^2 |U_{e 3}|^2 \sin^2{\left(\frac{\Delta m_{31}^2 L}{4E_\nu}\right)}\,,
\end{eqnarray}
where $|U_{\mu 3}|$ and $|U_{e 3}|$ are elements of the leptonic mixing matrix. In the standard parameterization~\cite{Tanabashi:2018oca}, $|U_{\mu 3}|^2 = \sin^2\theta_{23} \cos^2\theta_{13}$ and $|U_{e3}|^2 = \sin^2\theta_{13}$. We see here that measurements of the $K$DAR flux oscillating from $\nu_\mu$ to $\nu_e$ (or also measuring the survival probability $P(\nu_\mu \to \nu_\mu)$) will be sensitive predominantly to the mass-squared splitting\footnote{Given the approximate oscillation probability of interest, we are not sensitive to the sign of $\Delta m_{31}^2$, the so-called neutrino mass ordering. Going forward, we will assume $\Delta m_{31}^2 > 0$.} $\Delta m_{31}^2$ and the two mixing angles $\theta_{13}$ and $\theta_{23}$. Because $\sin^2\theta_{13}$ is small, the expected oscillation probability $P(\nu_\mu \to \nu_\mu)$ is much larger than $P(\nu_\mu \to \nu_e)$, and we will not expect very much sensitivity to $\theta_{13}$ at all.

Current measurements constrain $\Delta m_{31}^2 = \left(2.525^{+0.033}_{-0.031}\right) \times 10^{-3}$ eV$^2$, $\sin^2\theta_{13} = 0.02240^{+0.00065}_{-0.00066}$, and $\sin^2\theta_{23} = 0.582^{+0.015}_{-0.019}$~\cite{Esteban:2018azc}. In what follows, we will show sensitivity to these parameters using the $K$DAR flux measured at $295$ km. No parameter will be measured more precisely than next-generation (or current, for that matter) experimental constraints, however, this provides a consistency check on our understanding of the three-massive-neutrinos paradigm at a previously-unexplored combination of baseline length and neutrino energy.

\section{JSNS as a Kaon Decay-at-Rest Source for Hyper-Kamiokande}\label{sec:ExpDetails}
The J-PARC Spallation Neutron Source (JSNS) consists of a 3 GeV proton beam impinging on a mercury target. The target absorbs negatively-charged mesons, and positively-charged ones, specifically $\pi^+$ and $K^+$ will stop and decay at rest. JSNS$^2$ intends on measuring the predominantly muon-neutrino flux from the $\pi^+$ and $K^+$ decay-at-rest processes a distance of 24 m away from the target, as a means of searching for sterile-neutrino-induced oscillations with a new mass-squared splitting of $\Delta m_{41}^2 \approx 1$ eV$^2$~\cite{Ajimura:2017fld}. In addition to its purpose searching for sterile neutrinos, this detector can serve as a near detector for our proposed search, precisely constraining the $\nu_\mu$ flux and cross section for the $K$DAR neutrinos.
 
Currently in operation, the Super-Kamiokande (SK) detector is a distance of $L = 295$ km from J-PARC, and the proposed Hyper-Kamiokande (HK) detector will be at a similar distance. Even though the DAR flux is isotropic and a distance of $295$ km implies a large suppression to the flux, we will show that a handful of interactions from JSNS DAR neutrinos could be observed in HK. 

Here we estimate the number of $K$DAR events\footnote{The $\pi$DAR flux is too low energy to produce a large signal in Hyper-Kamiokande. We discuss this flux in the next section.} that can be observed a distance $D$ from J-PARC, assuming a detector with mass $M_\mathrm{Det.}$.  We assume that the JSNS will deliver $4\times 10^{22}$ protons on target (POT) per year of operation (corresponding to a beam power $P =1$ MW), and, per POT, approximately $10^{-2}$ $K^+$ are produced\footnote{Monte Carlo simulations indicate that $\sim 6\times 10^{-3}{-}1.1\times 10^{-2}$ $K^+$ are produced per POT when 3 GeV protons strike the J-PARC target~\cite{Jordan:2019xxx}.}. Accounting for the branching ratio $\mathrm{Br}(K^+ \to \mu^+ \nu_\mu) = 0.6356$~\cite{Tanabashi:2018oca}, this corresponds to $8 \times 10^{20}$ $K$DAR $\nu_\mu$ assuming $T = 3$ years of JSNS operation\footnote{JSNS is currently planned to operate beginning late 2019 for three years. Here, we imagine a hypothetical future run when the larger Hyper-Kamiokande detector is in operation.}. Because the DAR flux is isotropic, we may calculate it at SK/HK by scaling the produced rate by $1/4\pi D^2$, where $D \approx 295$ km is the distance from the JSNS source to Kamiokande:
\begin{equation}
\Phi^{K\mathrm{DAR}}_\mu = \frac{N_{\nu_\mu}}{4\pi D^2} = 7\times 10^4 \frac{\nu}{\mathrm{cm}^2} \left(\frac{T}{3\ \mathrm{yr}}\right) \left(\frac{P}{1\ \mathrm{MW}}\right).
\end{equation}

The flux of $\nu_\mu$, which can also oscillate between JSNS and Kamiokande, may interact with the water in SK or HK via charged current interactions. From Ref.~\cite{Formaggio:2013kya}, at $E_\nu = 236$ MeV,  this cross section is $\sigma_\mathrm{CCQE} \approx 1.7\times 10^{-39}$ cm$^2$ per nucleon. The total number of unoscillated events in a detector with mass $M_\mathrm{Det.}$ then is
\begin{eqnarray}
N^\mathrm{evt.}_\alpha &=&  \Phi^{K\mathrm{DAR}}_\mu \sigma_\mathrm{CCQE} N_\mathrm{nucleons}P(\nu_\mu \to \nu_\alpha), \\
&\approx& 30 \left(\frac{M_\mathrm{Det.}}{400\ \mathrm{kt}}\right) \left(\frac{T}{3\ \mathrm{yr}}\right) \left(\frac{P}{1\ \mathrm{MW}}\right) P(\nu_\mu \to \nu_\alpha).\ \
\end{eqnarray}
For clarity, we define the total exposure of the experiment to be $M_\mathrm{Det.} T P$. For our hypothetical future JSNS run concurrent with HK, we take $1200$ kt-MW-yr as our benchmark exposure. We also compare against a hypothetical $3600$ kt-MW-yr exposure, driven by a more powerful beam and/or longer data collection time.

In this estimation, we have ignored the effect of oscillations: if neutrinos oscillate according to the three-massive-neutrinos paradigm, with mixing angles and mass-squared splittings consistent with the most recent global fits of NuFit~\cite{Esteban:2018azc}, the oscillation probability at this energy and distance is $P(\nu_\mu \to \nu_\mu) \approx 0.50$ and $P(\nu_\mu \to \nu_e) \approx 0.04$. This implies that, for three years of JSNS operation and a fiducial volume of 400 kt, we can expect 15 $\mu^-$ charged current events and roughly 1 $e^-$ event. These event rates are sensitive, as discussed in Section~\ref{sec:Oscillations}, predominantly to the parameters $\theta_{23}$ and $\Delta m_{31}^2$. Using the measurement of a number of events at HK to infer a value of $P(\nu_\mu \to \nu_\mu)$ or $P(\nu_\mu \to \nu_e)$ requires a measurement of the flux at effectively zero baseline, which will be performed by the JSNS$^2$ experiment.

\subsection{Signal Identification and Background Reduction}
In this subsection we discuss the strategy for identifying the DAR neutrino events from JSNS, and how to reduce any possible background events. Assuming that this operation is concurrent with the planned T2HK (Tokai to Hyper-Kamiokande) beam-based experiment~\cite{Abe:2011ts}, the timing structure of the JSNS and T2HK beams will prevent any confusion between the two sources. The only remaining sources are cosmic ray muons and events from atmospheric neutrinos interacting in the fiducial volume of the detector.

The JSNS proton beam structure is as follows: two 80 ns bunches are delivered to the target, separated by 540 ns. This repeats at a rate of 25 Hz. Because the pion and kaon are both short-lived, the neutrinos coming from $\pi$DAR and $K$DAR will come promptly, nearly all inside of the two 80 ns bunches. This means that there is a window of $160$ ns with respect to the beam spacing of $4 \times 10^{4}$ $\mu$s in which to expect the DAR neutrinos at detection. Consequently, this means a reduction of backgrounds by a factor of $2.5 \times 10^{5}$.

Additionally, from the perspective of the detector, the direction of JSNS is well-known. Even if an outgoing muon or electron from a charged-current quasielastic interaction (a) does not go in the same direction as the incoming neutrino and (b) cannot be measured perfectly, this directionality helps in both observing the signal and reducing possible background events. The outgoing lepton direction is not the same as the incident neutrino direction, however, since the incoming neutrinos for the signal are monoenergetic, the outgoing lepton energy and direction are highly correlated. If the incoming neutrino direction is assumed, then, the reconstructed energy for these signal events will match the $\pi$DAR or $K$DAR values. This directionality and energy reconstruction will further reduce the expected background events significantly.

Given the combination these three handles (timing, direction, and reconstructed energy), the expected background for such searches is effectively zero.

\subsection{Results}\label{sec:Results}
We perform our analyses by constructing a Poissonian log-likelihood of the number of observed $\nu_\mu$ and $\nu_e$ CCQE events, respectively:
\begin{equation}\label{eq:Likelihood}
\mathcal{L}_\alpha = -\lambda_\alpha + x_\alpha\log{\lambda_\alpha} - \log{\left(x_\alpha!\right)},
\end{equation}
where $\alpha = e,$ $\mu$; $x_\alpha$ is the expected true number of events of flavor $\alpha$; and $\lambda_\alpha$ is the test number of events. We then use the test statistic $\Delta \chi^2 = -2\Delta(\mathcal{L}_e + \mathcal{L}_\mu)$. Without assuming anything about the structure of $P(\nu_\mu \to \nu_\mu)$ and $P(\nu_\mu \to \nu_e)$, we may ask how well we can constrain the individual probabilities at the baseline length/neutrino energy of interest here. As stated above, assuming the current best-fit values of the three-massive-neutrinos paradigm~\cite{Esteban:2018azc} as input parameters, $P(\nu_\mu \to \nu_\mu) \approx 0.5$ and $P(\nu_\mu \to \nu_e) \approx 0.04$. Using the likelihood in Eq.~(\ref{eq:Likelihood}), we may constrain at 68.3\% CL,
\begin{eqnarray}
&&P(\nu_\mu \to \nu_\mu) \in [0.38, 0.64], \\
&&P(\nu_\mu \to \nu_e) \in [0.016, 0.094].
\end{eqnarray}
Alternatively, observing the expected number of $K$DAR events in HK would lead to a  3$\sigma$ measurement of muon neutrino disappearance, $P(\nu_\mu\to\nu_\mu)\neq1$.

\begin{figure}
\centering
\includegraphics[width=\linewidth]{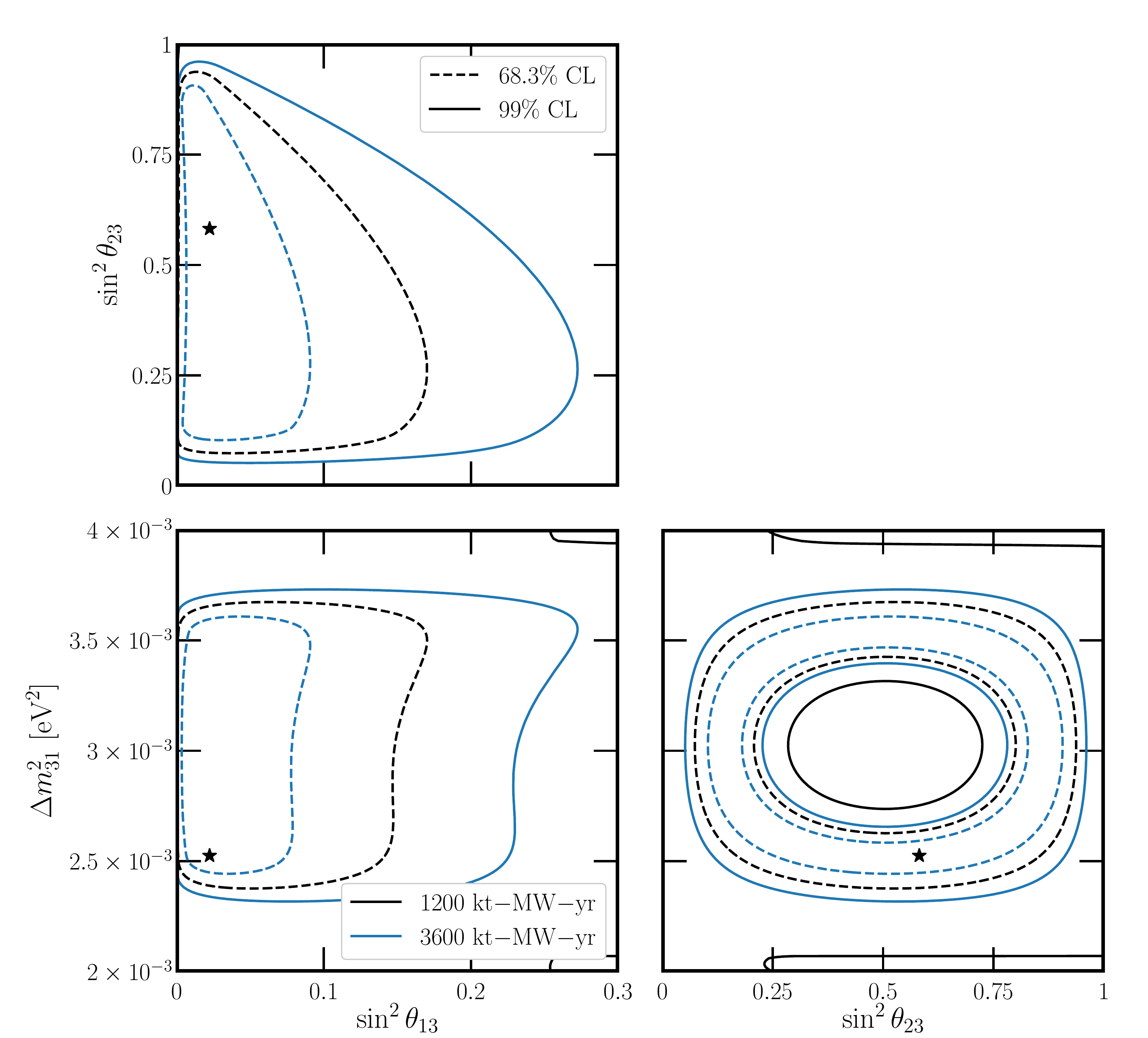}
\caption{Joint measurement capability of the $K$DAR neutrino events for the three neutrino oscillation parameters $\sin^2\theta_{13}$, $\sin^2\theta_{23}$, and $\Delta m_{31}^2$. Black lines are for an exposure of 1200 kt-MW-yr, and blue lines are an exposure of 3600 kt-MW-yr. The dashed lines show 68.3\% CL measurements, and the solid ones show 99\% CL. In each panel, the unseen third parameter has been marginalized over in our fit. Stars in each panel represent the assumed true value, consistent with the current best-fit values of the parameters.\label{fig:ThreeParamsNoPriors}}
\end{figure}
\begin{figure*}
\centering
\includegraphics[width=\linewidth]{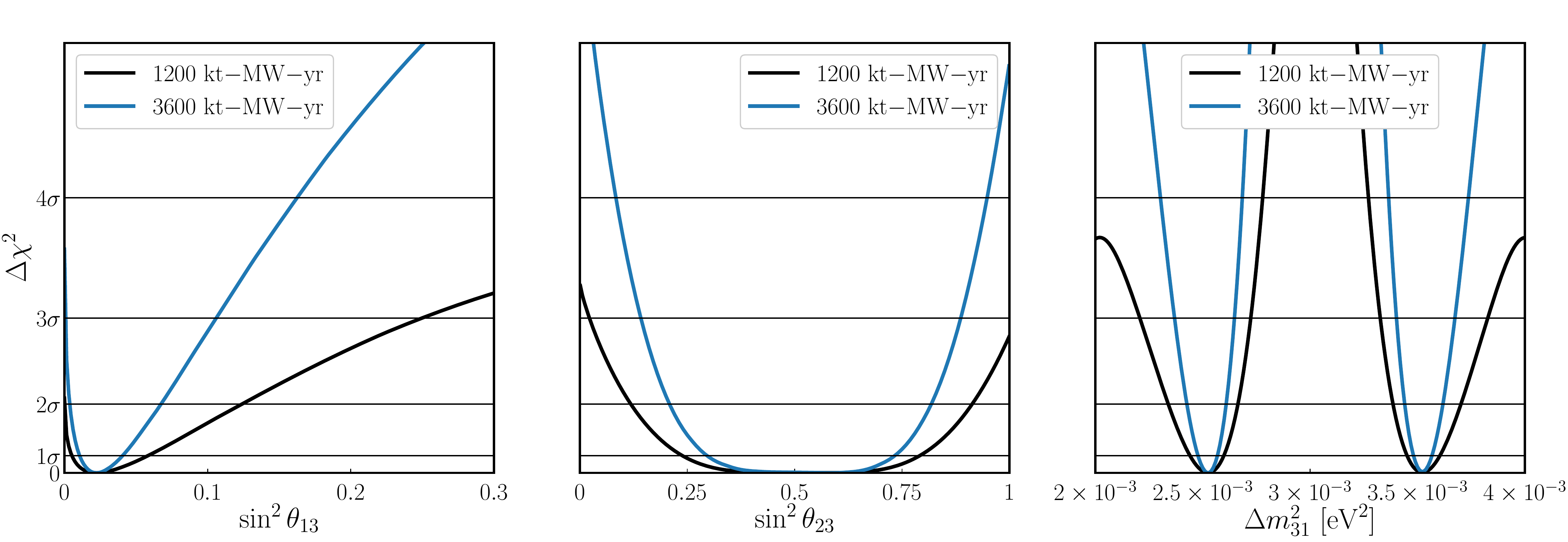}
\caption{Individual sensitivity to each neutrino oscillation parameter after marginalizing over the other two, incorporating priors on their values as discussed in the text. In each panel, the black line represents the measurement capability for an exposure of 1200 kt-MW-yr, while the blue line represents an exposure of 3600 kt-MW-yr. \label{fig:IndividualParams}}
\end{figure*}
We may go beyond this and ask, assuming the three-massive-neutrinos paradigm is correct, how well can we measure the relevant parameters, that is, $\sin^2\theta_{13}$, $\sin^2\theta_{23}$, and $\Delta m_{31}^2$. First, we attempt to constrain all three parameters without any outside information -- the ability to constrain these three parameters is shown in Fig.~\ref{fig:ThreeParamsNoPriors}. In Fig.~\ref{fig:ThreeParamsNoPriors}, we compare this result for the nominal exposure of 1200 kt-MW-yr in black with the optimistic exposure of 3600 kt-MW-yr in blue. For both exposures, we show the measurement capability at 68.3\% CL (dashed lines) and at 99\% CL (solid lines). The circular structure in the $\sin^2\theta_{23}$ vs. $\Delta m_{31}^2$ panel (bottom right) of Fig.~\ref{fig:ThreeParamsNoPriors} can be understood given the expression we discuss for $P(\nu_\mu\to \nu_\mu)$ in Eq.~(\ref{eq:Pmumu}), and persists even if we fix $\sin^2\theta_{13}$ to its best-fit value. Additionally, the pattern repeats for smaller and larger values of $\Delta m_{31}^2$ -- this is because we are measuring the oscillation probability for one value of $L$ and $E_\nu$, and smaller/larger $\Delta m_{31}^2$ values cause the oscillation to pass through the measured point. This can be understood by comparing this result with Fig.~\ref{fig:Probs}.

Lastly, we attempt to incorporate the existing knowledge of these three parameters, and analyze how well each of the three can be measured, including that existing knowledge on the other two as a prior on our analysis. For concreteness, we use $\Delta m_{31}^2 = \left(2.525^{+0.033}_{-0.031}\right) \times 10^{-3}$ eV$^2$, $\sin^2\theta_{13} = 0.02240^{+0.00065}_{-0.00066}$, and $\sin^2\theta_{23} = 0.582^{+0.015}_{-0.019}$~\cite{Esteban:2018azc}. The result of this procedure is shown in Fig.~\ref{fig:IndividualParams}. Unsurprisingly, no individual parameter is measured to more precision than is currently known, however, this is a measurement of these parameters at a new region in neutrino energy and baseline length, and serves as a consistency check on the three-massive-neutrinos paradigm.

\section{Other DAR opportunities}\label{sec:other}

In the previous section we pointed out that there is a new opportunity to further understand neutrino oscillations using KDAR with an source+detector pair that is part of the existing and planned facilities. In this section we discuss further opportunities to use DAR of muons and pions to test the three flavor oscillation picture. Though these opportunities are within the realm of possibility in terms of available source and detector technologies, there is not a concrete plan for the appropriate pair with the right baseline $L$ -- these possibilities would require new facilities, detectors or sources, but are worth exploring regardless. 

DAR neutrino sources that are not yet used for oscillation studies are already in existence or planning phases. In addition to J-PARC, which was discussed in the previous section, the Fermilab beam facility is itself also a source of DAR neutrinos, originating both in the 120~GeV NuMI beam~\cite{Grant:2015jva} and the 8~GeV BNB beam~\cite{Brice:2013fwa}. DAE$\delta$ALUS-like proton cyclotrons are also an interesting source which may be flexible in location and low in cost~\cite{Conrad:2009mh}. Measuring the interaction of DAR neutrinos in liquid Argon has been proposed by the CAPTAIN experiment~\cite{Berns:2013usa}. 
For the purpose of this section we will consider a source of pions decaying at rest (and subsequent muons decaying as well) at the order of $10^{23}$ decays per year. For detectors, we will consider large water, liquid Argon, and scintillator detectors of the scales of Hyper-K, DUNE, and JUNO respectively.

We will now briefly discuss the potential for $\pi$DAR neutrinos interacting in DUNE far detector, which would require a new $\pi$DAR source. We will also consider the oscillated $\mu$DAR neutrinos which are produced in facilities such as JSNS, the detection of which can be done by a new liquid scintillator detector similar to JUNO. 

\subsection{$\pi$DAR Neutrinos in a Liquid Argon Detector}

Here we consider the possibility of detecting the atmospheric frequency oscillation of $\pi$DAR neutrinos. Here $\nu_\mu$ CC scattering is below threshold, but there may be an opportunity to measure the $\nu_\mu\to\nu_e$ oscillation at low energy. This is difficult for water detectors because,
as mentioned above, the cross section of $\nu_e$ scattering on Oxygen has a threshold of roughly $11$ MeV, and even for $E_\nu = 30$ MeV, this cross section is still small~\cite{Kolbe:2002gk}. The cross section off of Hydrogen is larger, but the effective mass of Hyper-K in Hydrogen is about one eighth the total mass. 

Here we will consider $\nu_\mu\to\nu_e$ oscilation of $\pi$DAR neutrinos with a liquid Argon detector such as DUNE. The cross section for $\nu_e + \mathrm{Ar} \to e + \mathrm{K}$ for $E_\nu = 30$ MeV is approximately $2.5\times 10^{-40}$ cm$^2$~\cite{GilBotella:2003sz} per argon nucleus, several times larger than the CCQE rate in water. Focusing on the large DUNE far detector, performing this measurement would require a new source of stopped pions.

\begin{figure}
\centering
\includegraphics[width=\linewidth]{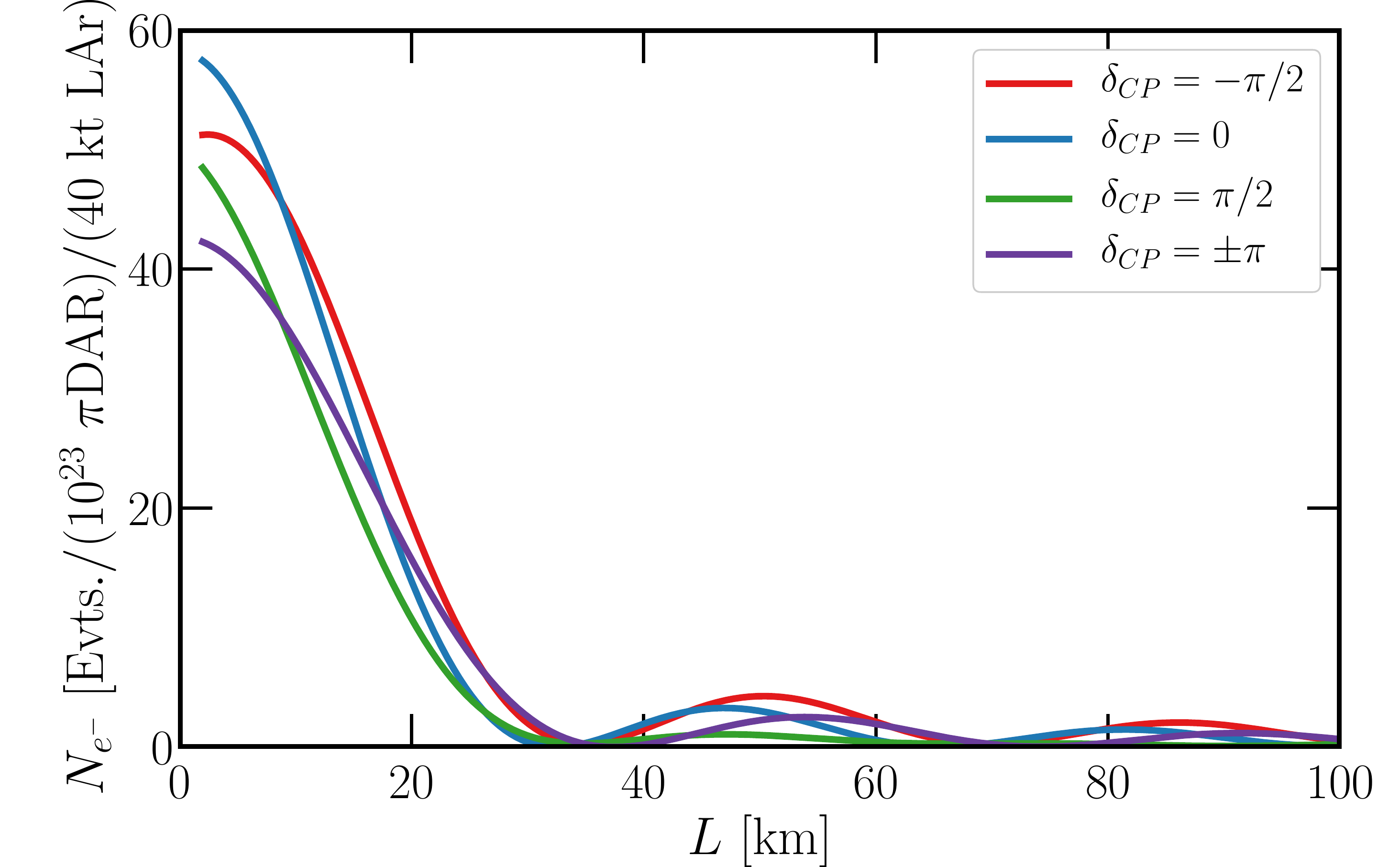}
\caption{Number of electron-neutrino events from pion decay-at-rest neutrinos that have undergone the oscillation $\nu_\mu \to \nu_e$ and interacted in a liquid argon detector. We present the number of events as a function of distance between the pion decay-at-rest source and the detector $L$, assuming a 40 kt liquid argon detector and a total of $10^{23}$ pion decays.\label{fig:PiDARDUNE}}
\end{figure}
Fig.~\ref{fig:PiDARDUNE} shows the number of such events as a function of the source-detector distance, assuming $10^{23}$ total pion decays and a 40 kiloton liquid Argon detector. We show the total number of events for four different choices of $\delta_\mathrm{CP}$, $-\pi/2$ (red), $0$ (blue), $\pi/2$ (green), and $\pm\pi$ (purple). Measuring this process would provide complementary information on the value of $\delta_\mathrm{CP}$ in addition to beam-~\cite{Acciarri:2015uup} and atmospheric-~\cite{Kelly:2019itm} based searches in the future.
 
\subsection{Subsequent Muon Decay-At-Rest Neutrino Spectrum}\label{subsec:MuDAR}

We have yet to discuss one remaining flux of neutrinos from JSNS: the neutrinos coming from muon decay-at-rest ($\mu$DAR), originating when the $\mu^+$, coming from the $\pi^+ \to \mu^+ \nu_\mu$ decay, stops and decays at rest. This produces a flux of $\overline{\nu}_\mu$ and $\nu_e$, each with a well-understood spectrum. The $\nu_e$ here, as with the oscillated $\nu_\mu$ from $\pi$DAR, are too low-energy to be of interest for scattering in large water detectors. On the other hand, the $\overline{\nu}_\mu$, which can oscillate into $\overline{\nu}_e$, may interact via the inverse beta decay reaction, which has a significantly larger cross section~\cite{Vogel:1999zy,Ankowski:2016oyj}. This cross section can be large for $\overline{\nu}_e p \to e^+ n$, scattering off the protons in Hydrogen, however, the inverse beta decay cross section for scattering off Oxygen is too small to be of use~\cite{Kolbe:2002gk}.

Limiting ourselves to the JSNS to HK setup considered in Section~\ref{sec:ExpDetails}, with a large water detector, our realistic exposure assumption of $1200$ kt-MW-yr predicts roughly $1$ $\overline{\nu}_\mu \to \overline{\nu}_e$ event. For the long exposure of $3600$ kt-MW-yr, we expect $\sim 3-5$ events, where the number of events depends significantly on the mixing parameters, especially $\delta_{CP}$. 
\begin{figure}
\centering
\includegraphics[width=\linewidth]{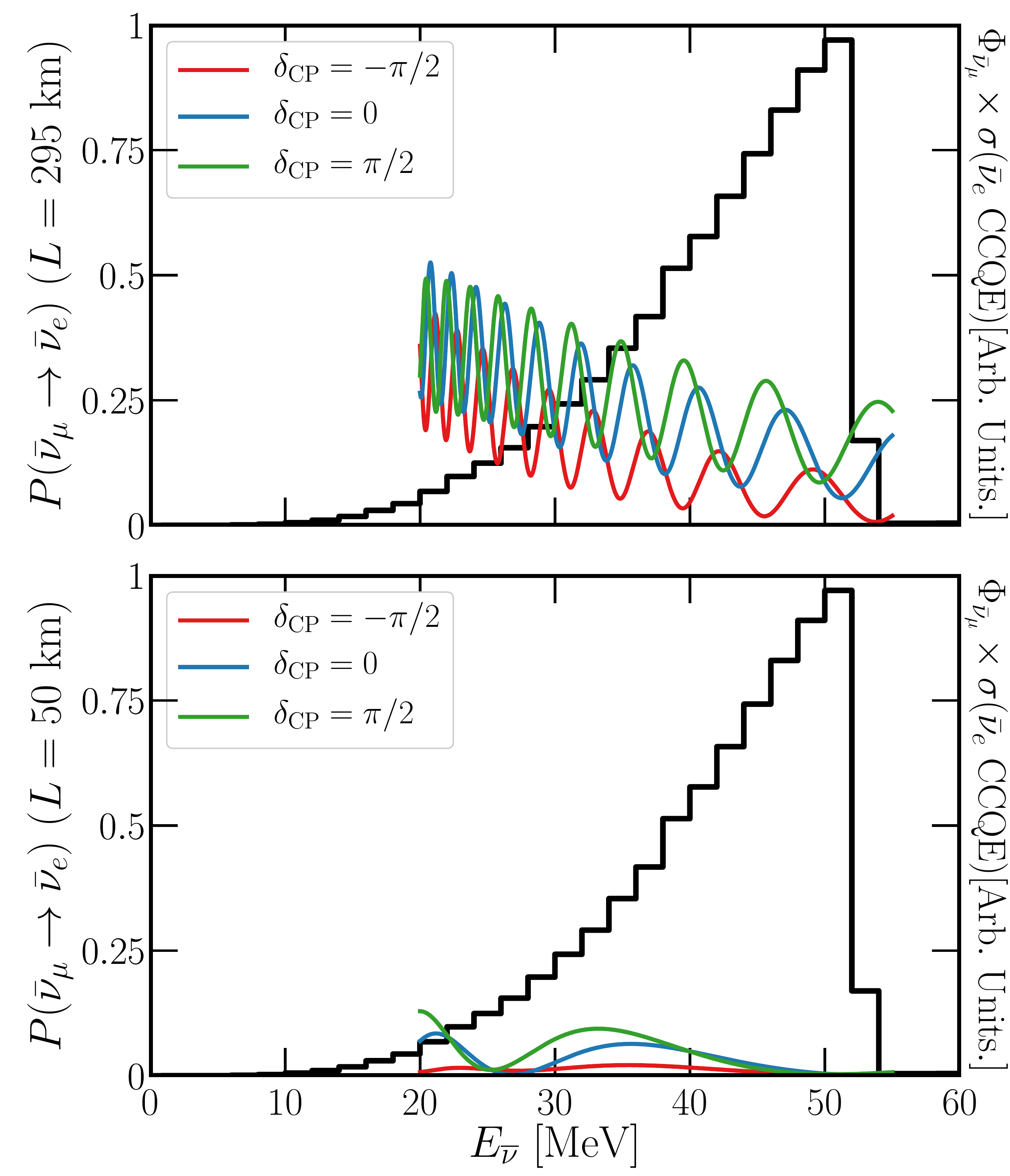}
\caption{Black: The expected shape (flux times cross section) of $\overline{\nu}_\mu$ emerging from $\mu$DAR, oscillating into $\overline{\nu}_e$, and interacting via inverse beta decay reactions. Colored lines: oscillation probabilities for this process, assuming $\delta_\mathrm{CP} = -\pi/2$ (red), $0$ (green), or $\pi/2$ (purple). Top panel shows oscillation probabilities for $L = 295$ km, bottom shows for $L = 50$ km.
\label{fig:muDARSpectrum}}
\end{figure}

If a large detector, perhaps consisting of liquid scintillator like JUNO~\cite{An:2015jdp}, were constructed near a $\pi$DAR source, this would lead to a large number of $\overline{\nu}_e$ charged current events coming from $\mu$DAR. Additionally, neutron reconstruction techniques, such as gadolinium loading~\cite{Beacom:2003nk}, could assist in identifying these events and measuring them precisely. Fig.~\ref{fig:muDARSpectrum} (black) shows the expected $\overline{\nu}_\mu$ flux convolved with the $\overline{\nu}_e$ inverse beta decay cross section that could be measured in such a scenario.

Without focusing on a specific detector or concrete numbers of expected events, we may also consider what the oscillation probability $P(\overline{\nu}_\mu \to \overline{\nu}_e)$ is for these energies. As evident in Fig.~\ref{fig:Probs}, oscillations are fast for these energies if $L \gtrsim 50$ km -- we overlay this oscillation probability for $L = 295$ km (the distance from JSNS to HK) in the top panel and for $L = 50$ km in the bottom panel of Fig.~\ref{fig:muDARSpectrum}. For each panel, we show three curves for the oscillation probability: $\delta_\mathrm{CP} = -\pi/2$ (red), $0$ (blue), and $\pi/2$ (green). While such a measurement would likely be weaker than the upcoming DUNE or HK beam-based measurements, it provides a complementary approach to measuring CP-violation in the lepton sector.

\section{Discussion \& Conclusions}\label{sec:Conclusions}
In this work, we have demonstrated that the future Hyper-Kamiokande experiment will have sensitivity to measure neutrino events originated from decay-at-rest processes in the J-PARC Spallation Neutron Source experiment. Because of the sharp features of this event sample, coming from kaon decay-at-rest, we may be able to further our knowledge of neutrino oscillations for a specific value of neutrino energy and baseline distance. This region of neutrino energy and baseline distance occupies a space unexplored, to date. This would constitute the first time that decay-at-rest neutrinos will be detected after having undergone oscillations.

We have explored how this setup is capable of measuring the oscillation probabilities $P(\nu_\mu \to \nu_\mu)$ and $P(\nu_\mu \to \nu_e)$, as well as certain oscillation parameters, specifically $\sin^2\theta_{13}$, $\sin^2\theta_{23}$, and $\Delta m_{31}^2$. While these measurements will not be as powerful as experiments running concurrently with this proposal, the added information (namely because it is in a very different baseline and energy region and for a fixed neutrino energy) contributes significantly to a better understanding of neutrino oscillations.

Additionally, other decay-at-rest neutrino sources, specifically those from decays of stopped pions and muons, can provide even further insight into neutrino oscillations at lower energies. In this work, we explored the challenges of measuring these neutrinos at long distance, and also showed that measuring them would provide further information on the three-massive-neutrinos paradigm with respect to CP-violation in the lepton sector.

While there is no current plan to operate JSNS concurrently with Hyper-Kamiokande, the physics cases of simultaneous operation have yet to be fully explored. This work demonstrates a small but important subset of the potential of measuring neutrino oscillations from decay-at-rest processes at long baselines. With the next generation of neutrino experiments planning to make precision measurements of the remaining unknown oscillation parameters, the neutrino physics community should focus on over-constraining the three-massive-neutrinos paradigm, and measurements like this across a wide range of baseline lengths and neutrino energies are crucial for such tests.

\acknowledgments
We thank Andr{\'e} de Gouv{\^e}a, Johnathon Jordan, and Joshua Spitz for valuable discussions regarding this work. Fermilab is operated by the Fermi Research Alliance, LLC under contract No. DE-AC02-07CH11359 with the United States Department of Energy.


\bibliographystyle{JHEPEsque}
\bibliography{references}

\end{document}